\documentclass{article}
\usepackage[affil-it]{authblk}
\usepackage{amsmath, amsfonts, amsthm, amssymb}
\usepackage[usenames,dvipsnames]{color}
\usepackage{comment}
\usepackage{ifpdf}
\usepackage{setspace}
\usepackage[utf8]{inputenc}
\usepackage[english]{babel}
\usepackage{graphicx}  
\usepackage{breqn}
\usepackage{mathrsfs}
\usepackage{gensymb}
\usepackage{caption}
\usepackage{pgfplots}
\usepackage{cite}
\usepackage[left=.8in,right=.8in,top=.8in,bottom=.8in]{geometry}  
\numberwithin{equation}{section}
\newcommand*\colvec[3][]{
    \begin{pmatrix}\ifx\relax#1\relax\else#1\\\fi#2\\#3\end{pmatrix}
}

\begin{document}
\title{Where are the BTZ Black Hole Degrees of Freedom?}   
\author{Joseph M. Mitchell}
\affil{Department of Physics, University of California, Davis}

\date{\today}      
\maketitle

\begin{abstract}
\noindent Previous derivations of the BTZ black hole entropy from a dual conformal description place the degrees of freedom at spatial infinity. Here it is shown for the non-rotating case that a dual conformal description exists at any location around the black hole, a result that has a strong physical appeal considering that in 2+1 dimensions there are no propagating degrees of freedom in the classical theory. Two copies of the central charge of $3\ell/2G$ are recovered, and the microcanonical Cardy formula yields the correct Bekenstein-Hawking entropy. 
\end{abstract}

\pagebreak 

\section{Introduction}
Black holes were first shown to be thermodynamic objects with temperatures and entropies by Hawking \cite{Hawk} and Bekenstein \cite{Beken} more than 40 years ago. However, the exact nature of the states that account for the degrees of freedom of the black hole has remained mysterious. One might think that without a full candidate theory of quantum gravity it would be difficult to say anything at all about the details of the states. Nevertheless, there now exist many seemingly different microscopic descriptions that all give the correct entropy. A proposed solution to this unexpected universality is to account for the degrees of freedom through a classical symmetry that is independent of the actual details of the quantum gravity except that such a quantum theory must exist. A two dimensional dual conformal description is a candidate for this approach. The idea was first suggested in \cite{ConformOrigin} and then fully implemented by Strominger \cite{Strominger} (and independently by Birmingham et al. \cite{Birmingham_CFT}) for the BTZ black hole \cite{BTZsolution} by applying Brown-Henneaux boundary conditions \cite{BrownADS}.

Despite the elegance of Strominger's derivation, there remain some conceptual hurdles. In particular, Brown and Henneaux impose boundary conditions at spatial infinity, which means the dual CFT lives on the asymptotic boundary of the BTZ spacetime. A physically more appealing location for the degrees of freedom, especially in higher dimensional cases, is the event horizon of the black hole \cite{CarlipBCHorizon, SolodukhinBCHorizon, CarlipBCHorizon2}. For example, if there exists a dual CFT living on the event horizon this would be a step towards understanding why the entropy scales with area instead of volume. However, in 2+1 dimensions this still seems somewhat insufficient considering there are no local degrees of freedom in the classical theory. This is troublesome when one considers that the goal of this approach is to account for the black hole entropy through a classical symmetry. Another source of ambiguity in the dual CFT approach is in the choice of boundary conditions. The most common procedure is to fix the intrinsic geometry of the boundary, but one could just as well impose restrictions on the extrinsic curvature. For example, in \cite{Whatwedontknow} Carlip shows that in the Chern-Simons formulation one can derive the correct entropy of the BTZ black hole by imposing various different choices of boundary conditions, which include fixing the induced metric on the boundary, the mean extrinsic curvature and York's extrinsic time. It almost seems as though, at least in the Chern-Simons formulation, the choice of boundary conditions does not matter, which adds another layer of mystery to the approach. 

The main line of this paper is to take some steps towards resolving the conceptual difficulties described above for the case of the non-rotating BTZ black hole. In particular, through a consistent set of boundary conditions, a dual CFT is found to exist not just on the event horizon or spatial infinity but at any location around the black hole. Two copies of the central charge $c^{\pm} = 3\ell/2G$ are recovered and the microcanonical version of the Cardy formula is used to compute the correct Bekenstein-Hawking entropy. 


\section{A Dual CFT at Any Location}
Let us begin with the non-rotating BTZ black hole \cite{BTZsolution} written in the familiar Schwarzschild-like coordinates

\begin{equation}
	ds^2 = - (r^2 - r_+^2) dt^2 + \frac{ \ell^2 dr^2}{r^2 - r_+^2} + r^2 d\phi^2
\end{equation}	
where the cosmological constant $\Lambda = - 1/\ell^2$, $r_+^2 = 8GM\ell^2$ is the location of the event horizon, $M$ is the mass of the black hole and I have chosen a dimensionless time coordinate $t \rightarrow t/\ell$. The diffeomorphisms preserving the asymptotic structure of (2.1) have functional dependence on the combinations $t \pm \phi$  \cite{BrownADS}, which are null at spatial infinity and correspond to left- and right-moving modes on the asymptotic boundary. Therefore, as a useful starting point, the transformation below is used to put the metric (2.1) into a form adaptable to circular null coordinates at any location:
\begin{equation}
	\begin{split}
		r &= r_+ e^{\rho/\ell - r_+ \tau / \ell} \\
		t & = \tau - \frac{\ell}{2r_+}\ln\Big(1 - e^{-2\rho / \ell + 2r_+ \tau / \ell}\Big) \\
	\end{split}
\end{equation}	
The metric now takes the simple form

\begin{equation}
	ds^2 = r_+^2 e^{2\chi}(- d\tau^2 + d\phi^2) + d\rho^2
\end{equation}	
where $\chi =  \rho/\ell - r_+ \tau/\ell$ and the combination $\tau \pm \phi$ is clearly null on any surface of constant $\rho$. Some properties of an observer in the $(\tau, \rho, \phi)$ coordinate system are discussed later. With the metric in this new form, one can consistently impose the following restrictions on diffeomorphisms of the intrinsic geometry:
\begin{equation}
	\begin{split}
		\mathcal{L}_{\xi} g_{\tau \tau} &= 0  \ \Rightarrow \ \xi^{\rho} = r_+ \xi^{\tau} - \ell \partial_{\tau}\xi^{\tau} \\
		\mathcal{L}_{\xi} g_{\phi \phi} &= 0  \ \Rightarrow \ \xi^{\rho} = r_+ \xi^{\tau} - \ell \partial_{\phi}\xi^{\phi} \\	
		\mathcal{L}_{\xi} g_{\phi \tau} &= 0 \  \Rightarrow \ \partial_{\phi}\xi^{\tau} = \partial_{\tau}\xi^{\phi} \\	
		\mathcal{L}_{\xi} g_{\rho \rho} &= 0 \ \Rightarrow \ \partial_{\rho} \xi^{\rho} = 0 
	\end{split}		
\end{equation}	
while allowing $\mathcal{O}(1)$ changes in $g_{\rho \tau}$ and $g_{\rho \phi}$. The above boundary conditions are preserved by diffeomorphisms of the form
\begin{equation}
	\begin{split}
		\xi^{\pm \tau} &= \frac{1}{2} T^{\pm}(\tau \pm \phi)\\
		\xi^{\pm \rho} &= \frac{r_+}{2} T^{\pm}(\tau \pm \phi) \mp \frac{\ell}{2} \partial_{\phi} T^{\pm}(\tau \pm \phi) \\
		\xi^{\pm \phi} &= \pm \frac{1}{2} T^{\pm}(\tau \pm \phi) 
	\end{split}	
\end{equation}
where $T^{\pm}$ depends only on the combination $\tau \pm \phi$ and is otherwise arbitrary. The diffeomorphisms (2.5) almost agree with the asymptotic symmetries of Brown and Henneaux to first order in $\rho$ as well as the vector field discussed in \cite{BoundaryDynamics}, but there are two key differences. First, the vector field here depends on a different time coordinate that approaches the usual Schwarzschild-like time only at spatial infinity. Second, the first term in the radial component above contains no derivatives on $T^{\pm}$ and is proportional to the size of the horizon.  
		
The factor of $1/2$ in (2.5) normalizes the vector field such that its commutator is

\begin{equation}
	\begin{split}
		[\xi^{\pm}(T^{\pm}_1), \xi^{\pm}(T^{\pm}_2)] &= \xi^{\pm}(T^{\pm}_1\partial_{\phi}T^{\pm}_2 - T^{\pm}_2\partial_{\phi}T^{\pm}_1)\\
		[\xi^{+}(T^{+}_1), \xi^{-}(T^{-}_2)] &= 0
	\end{split}
\end{equation}		
By expanding $T^{\pm}$ into modes $T_n^{\pm} = e^{in(\tau \pm \phi)}$ this can be cast into a form that is often more convenient and perhaps more familiar:

\begin{equation}
	\begin{split}
		[\xi^{\pm}_n, \xi^{\pm}_m] &= i(m-n)\xi^{\pm}_{n + m}\\
		[\xi^{+}_n, \xi^{-}_m] &= 0
	\end{split}
\end{equation}		
This is recognizable as a pair of commuting Witt Algebras, which suggests the applicability of techniques from two dimensional conformal field theory. Note that no subleading terms have been neglected and the Witt algebra holds exactly at all values of $\rho$.

In order to fully recover the elegant derivation of Strominger, however, one must search for nonzero central terms in the algebra of the generators. This is done most easily by switching to the canonical formalism, briefly reviewed in the Appendix. From the general ADM form of the metric

\begin{equation}
	ds^2 = -N^2 dt^2 + q_{ij}(dx^i + N^i dt)(dx^j + N^j dt)
\end{equation}
it is easy to read off the lapse $N$ and shift $N^i$ from (2.3):

\begin{equation}
	N = r_+e^{\rho/\ell - r_+\tau/\ell}, \ \ \ N^i = 0
\end{equation}
The only non-vanishing component of the momentum (A.2) canonically conjugate to the spatial metric $q_{ij}$ is

\begin{equation}
	\pi^{\rho \rho} = \frac{r_+}{\ell}
\end{equation}		
We must also replace spacetime diffeomorphisms $\xi^{\mu}$ with the corresponding surface deformation parameters $(\xi^{\perp}, \hat{\xi}^i)$, which are related through  \cite{TeitSDParameter}

\begin{equation}
	\xi^{\perp} = N\xi^t, \ \ \hat{\xi}^i = \xi^i + N^i \xi^t
\end{equation}

Here some care must be taken in transforming the boundary conditions (2.4) on the full spacetime metric $g_{\mu \nu}$ under Lie transport into boundary conditions on the canonical variables $(q_{ij}, \pi^{ij})$ under Hamiltonian transport. Neglecting boundary terms for the moment, the transformation generated under Hamiltonian transport (A.5) yields

\begin{equation}
	\begin{split}
		&\delta_{\xi} q_{\rho \rho} = 0 \ \Rightarrow \ \partial_{\rho} \hat{\xi}^{\rho} = 0 \\ 
		&\delta_{\xi} q_{\phi \phi} = 0 \ \Rightarrow \ \hat{\xi}^{\rho} = \frac{\ell \pi}{\sqrt{q}} \xi^{\perp} - \ell \partial_{\phi} \hat{\xi}^{\phi}
	\end{split}
\end{equation}	
The lapse $N$ and shift $N^i$ will still transform according to the Lie derivative (2.4), rewritten below in terms of ADM variables:
\begin{equation}
	\begin{split}
		\delta_{\xi} N \ &= 0 \ \Rightarrow \ N \hat{\xi}^{\rho} = -\ell \partial_{\tau} \xi^{\perp} \\
		\delta_{\xi} N^{\phi} &= 0 \ \Rightarrow \ \partial_{\tau}\hat{\xi}^{\phi} = N q^{\phi \phi}\partial_{\phi} \xi^{\perp} \\
	\end{split}
\end{equation}	
Allowed changes in $N^{\rho}$ and $q_{\rho \phi}$ are, of course, still $\mathcal{O}(1)$. Additionally, since we do not require variations in $q_{\rho \phi}$ vanish, we can also consistently fix the off-diagonal element of the ADM momentum

\begin{equation}
	\delta_{\xi} \pi^{\rho \phi} = 0 \Rightarrow \pi^{\rho \rho} \partial_{\rho} \hat{\xi}^{\phi} - \sqrt{q} q^{\rho \rho} q^{\phi \phi}\Big(\partial_{\rho} \partial_{\phi} -\frac{1}{\ell} \partial_{\phi}\Big) \xi^{\perp}
\end{equation}
The transformed boundary conditions (2.12)-(2.14) are preserved by surface deformations of the form
\begin{equation}
	\begin{split}
		\xi^{\pm \perp} &= \frac{r_+}{2}e^{\rho/\ell - r_+\tau/\ell} \ T^{\pm}(\tau \pm \phi) \\
		\hat{\xi}^{\pm \rho} &= \frac{r_+}{2} T^{\pm}(\tau \pm \phi) \mp \frac{\ell}{2} \partial_{\phi} T^{\pm}(\tau \pm \phi) \\
		\hat{\xi}^{\pm \phi} &= \pm \frac{1}{2} T^{\pm}(\tau \pm \phi) 
	\end{split}	
\end{equation}
From (2.9) and (2.11) it is easy to see that the surface deformations (2.15) correspond to the same diffeomorphisms (2.5) obtained by placing boundary conditions on the full spacetime metric. In other words, there is no need to adjust the set of allowed diffeomorphisms when switching to the canonical formalism and (2.14) has been obtained basically for free. For completeness I shall also mention that, for the transformations discussed here, variations in $\pi^{\rho \rho}$ are $\mathcal{O}(1)$ while changes in $\pi^{\phi \phi}$ appear to vanish. However, we do not impose that $\pi^{\phi \phi}$ be held fixed as a boundary condition since we are already holding $q_{\phi \phi}$ fixed. I speculate that the vanishing of variations in $\pi^{\phi \phi}$ might fall naturally out of some combination of the boundary conditions above for the particular case discussed here.

There is an important subtlety to notice from the second line of (2.12). Using the boundary conditions (2.13)-(2.14), one can tell that neither $\xi^{\perp}$ nor $\hat{\xi}^{\phi}$ depend on $\pi$ or $\sqrt{q}$. Then, from (2.12), it is clear that the radial component $\hat{\xi}^{\rho}$ is, in fact, dependent on the trace of the ADM momentum and the determinant of the spatial metric. This is important because, when the surface deformations depend on the canonical variables, they can have nontrivial Poisson brackets with the generators. This means one cannot use the simpler and more well known Lie bracket of surface deformations \cite{TeitSDParameter}:

\begin{equation}
	\begin{split} 
		\{\xi, \eta\}^{\perp}_{SD} &= \hat{\xi}^i D_i \eta^{\perp} - \hat{\eta}^i D_i \xi^{\perp} \\
		\{\xi,\eta\}^i_{SD} &= \hat{\xi}^k D_k \eta^i - \hat{\eta}^k D_k \hat{\xi}^i + q^{i k} (\xi^{\perp} D_k \eta^{\perp} - \eta^{\perp} D_k \xi^{\perp}) \\
	\end{split}	
\end{equation}
Instead, as discussed in the Appendix, one must use the full surface deformation bracket (A.9). Using the full Lie bracket for the case at hand we again obtain a pair of commuting Witt algebras:

\begin{equation}
	\begin{split}
		&\{\xi^{\pm}_n,\xi^{\pm}_m\}^{\perp} = i(m - n)\xi^{\pm \perp}_{n+m} \ \ \ \ \{\xi^{+}_n,\xi^{-}_m\}^{\perp} = 0 \\
		&\{\xi^{\pm}_n,\xi^{\pm}_m\}^i \ = i(m - n)\hat{\xi}^{\pm i}_{n+m} \ \ \ \  \{\xi^{+}_n,\xi^{-}_m\}^i \ = 0 \\
	\end{split}
\end{equation}		
where I have again used the mode expansions of $T^{\pm}$ for simplicity of notion. Note that had we naively used the bracket (2.16) instead of (A.9) then the radial component would have spoiled the above Witt algebra. 

Here, to be careful, I should mention a couple of observations. First, the dependence of $\hat{\xi}^{\rho}$ on the canonical variables does slightly change the transformation generated by the Hamiltonian by adding terms to (A.5) (and, therefore, (2.12) and (2.14)) proportional to the constraints. However, when evaluated on-shell, Hamiltonian transport will still generate a transformation equivalent to Lie transport and, even off-shell, this subtlety does not affect the Witt algebra above. Second, one might worry that the surface deformations could have additional dependence on the canonical variables that I have not mentioned. Indeed, this is actually true. In particular, $\xi^{\perp}$ depends on the lapse, which depends on $\rho$. From the metric (2.3) it is clear that on a constant time slice $\rho$ is the proper distance to the horizon and, therefore, a metric dependent quantity. However, this does not affect the algebra (2.17) because this adds to the Lie bracket terms proportional to $\{q_{\rho \rho}, H[\xi]\}$, which vanish under the boundary conditions considered here. So there should be no reason to worry. 

Finally, it is now time to search for the appearance of central charges in the algebra of the generators. Turning to the general form for the central terms in the canonical formalism (A.8) derived by Carlip \cite{Whatwedontknow}, it is easy to see that the relevant three derivative terms are

\begin{equation}
	K[\xi, \eta] = ... - \frac{1}{8\pi G} \int{ d\phi \sqrt{\sigma} n^k (D_m \hat{\xi}_k D^m \eta^{\perp} - D_m \hat{\eta}_k D^m \xi^{\perp})}
\end{equation}	
where $\sigma$ is the determinant of the induced metric on the boundary, $n^k$ is the unit normal and $D_m$ is the spatial covariant derivative compatible with the spatial metric $q_{ij}$. Evaluating the above expression with the diffeomorphisms (2.5) yields

\begin{equation}
	K[\xi^{\pm}(T_1^{\pm}), \xi^{\pm}(T_2^{\pm})] = ... \mp \frac{\ell}{32 \pi G} \int{d\phi (\partial_{\phi} T_1^{\pm}\partial_{\phi}^2 T_2^{\pm} - \partial_{\phi} T_2^{\pm}\partial_{\phi}^2 T_1^{\pm})} \\
\end{equation}	
which can be recognized as a central term in a Virasoro Algebra with central charges \cite{CFTyellowpages}

\begin{equation}
	c^{\pm} = \frac{3\ell}{2G}
\end{equation}
This agrees exactly with central charges obtained by Brown and Henneaux when studying the asymptotic symmetries of asymptotically AdS$_3$ spacetimes but holds at any spatial location. 

\section{The Conformal Weight}
In order to calculate the entropy of the black hole from the microcanonical version of the Cardy formula \cite{Cardy1, Cardy2}

\begin{equation}
	S = 2\pi \sqrt{\frac{c}{6}\Big(\Delta - \frac{c}{24}\Big)} 
\end{equation}
one needs to know the conformal weight $\Delta$. The conformal weight comes from the boundary term $B[\xi]$, which must be chosen to cancel the variation in the Hamiltonian \cite{HamBterms} in order to obtain a well defined variational principle. This only determines the boundary term up to a constant, which must be fixed through the physics. For the metric and boundary conditions at hand, the remaining terms in the variation of the Hamiltonian (A.10) are

\begin{equation}
	\delta H[\xi] = ... -\frac{1}{16\pi G} \int d\phi \Big\{ \sqrt{\sigma} \Big[\sigma^{\phi \phi} n^{\rho} \xi^{\perp} \Big(D_{\phi} \delta q_{\rho \phi} - D_{\rho} \delta q_{\phi \phi}\Big) - D_{\phi} \xi^{\perp} n^{\rho} \sigma^{\phi \phi} \delta q_{\rho \phi} \Big] + 2 \hat{\xi}^{\rho} \delta \pi^{\rho}_{\rho} \Big\}
\end{equation}	
The first two terms can be rewritten as the variation of the mean extrinsic curvature $k = \sigma^{\phi \phi} D_{\phi}n_{\phi}$ of the boundary as viewed as a submanifold. The third term can be rewritten in terms of the variation of the normal \newline $\delta n^k = -\frac{1}{2} q^{k i} n^j \delta q_{i j}$, which is obtained from $\delta (q_{i j} n^i n^j) = 0$. Some care must be taken when handling the last term because, as noted above, $\hat{\xi}^{\rho}$ is dependent on the canonical variables:

\begin{equation}
		\delta_{\xi} (\hat{\xi}^{\rho} \pi_{\rho}^{\rho}) = \hat{\xi}^{\rho} \delta\pi_{\rho}^{\rho} + \pi_{\rho}^{\rho} \frac{\ell}{\sqrt{q}} \xi^{\perp} \delta\pi
\end{equation}	
where the second term comes from the variation of $\hat{\xi}^{\rho}$ and use was made of (2.12). The boundary term can thus be written as

\begin{equation}
	B[\xi] = \frac{1}{8\pi G} \int{d\phi\Big[\sqrt{\sigma} \Big(n^m \partial_m \xi^{\perp} - k\xi^{\perp} \Big) + \hat{\xi}^{\rho} \pi^{\rho}_{\rho} - \frac{\ell}{2\sqrt{q}} \pi^2 \xi^{\perp} \Big]} + B_0
\end{equation}
where $B_0$ is the arbitrary constant mentioned above. Evaluating (3.4) with zero modes $T^{\pm}_0 = 1$ gives
\begin{equation}
	B[T^{\pm}_0]  = \frac{r_+^2}{16G\ell} + B_0
\end{equation}	
At this point it is now clear that we can fix the additive constant using the same method as Strominger. Since the non-rotating BTZ metric (2.1) becomes that of AdS$_3$ when $r_+^2 \rightarrow -\ell^2$, by requiring the conformal weight to vanish for AdS$_3$ one obtains
\begin{equation}
B[T^{\pm}_0] = \Delta^{\pm} = \frac{r_+^2}{16G\ell} + \frac{\ell}{16G}
\end{equation}
Using the above conformal weights and the central charges (2.20), the microcanonical Cardy formula then yields
\begin{equation}
	S = 2\pi \sqrt{\frac{c^+}{6}\Big(\Delta^+ - \frac{c^+}{24}\Big)} +  2\pi \sqrt{\frac{c^-}{6}\Big(\Delta^- - \frac{c^-}{24}\Big)}= \frac{2\pi r_+}{4G}
\end{equation}
which is the correct Bekenstein-Hawking entropy. As in the derivation of the central charges, the analysis here has the advantage of holding at any spatial location. 


\section{Observer on a Constant $\rho$ Surface}
As far as I am aware, the system of coordinates obtained under the transformation (2.2) has not been used before. It is therefore useful to understand what an observer on a constant $\rho$ surface will see in terms of more familiar coordinate systems. As a start, from the metric (2.3), it is easy to see that a stationary observer in the $(\tau, \rho, \phi)$ coordinate system has a four-velocity $u^{\mu} = (\frac{1}{r_+}e^{-\rho/\ell + r_+\tau/\ell}, 0, 0)$, which yields a proper acceleration of 

\begin{equation}
	a^{\mu} = u^{\nu} \nabla_{\nu} u^{\mu} = (0, 1/\ell, 0) \ \Rightarrow \ a^2 = g_{\mu \nu} a^{\mu} a^{\nu} = \frac{1}{\ell^2}
\end{equation}	
So an observer on a surface of constant $\rho$ will feel a constant radial acceleration proportional to the cosmological constant. In order to see this motion graphically, we start with the metric in terms of Schwarzschild-like coordinates (2.1) and then make a transformation to a tortoise-like coordinate 

\begin{equation}
r = r_+\coth(r_+r_{*} / \ell)
\end{equation}
The metric now takes the form

\begin{equation}
	ds^2 = \frac{r^2_+}{\sinh^2(r_+r_* / \ell)} (-dt^2 + dr^2_*) + r_+^2 \coth^2(r_+r_* / \ell) d\phi^2
\end{equation}
In terms of the tortoise-like coordinate, radial null geodesics are just straight lines at $45 \degree$ angles, just as in Minkowski space. Also note that, just as one would expect for a tortoise-like coordinate, the black hole horizon has been pushed to $r_{*} \rightarrow \infty$ and spatial infinity has been mapped to $r_{*} \rightarrow 0$. From (2.2), it is now easy to plot the motion of an observer on a constant $\rho$ surface in the $r_{*} - t$ plane:

\begin{center}
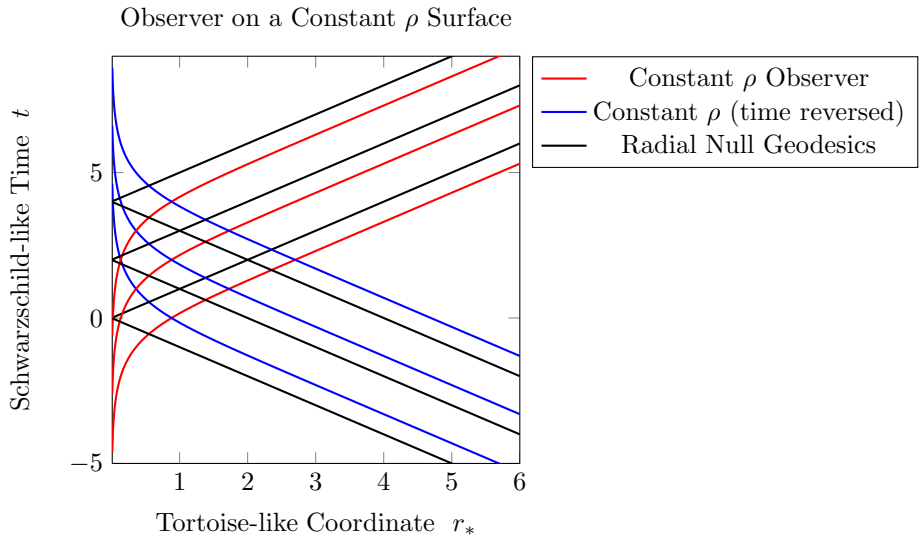

\begin{tikzpicture}
	\begin{axis}[
        xmin=0.01, xmax=6, 
       xticklabel style={/pgf/number format/.cd,fixed}, 
        width=7cm, height=7cm,
        ymin=-5,      
        ymax= 9,      
        axis background/.style={fill=white},
        ylabel={Schwarzschild-like Time \ $t$},
        xlabel={Tortoise-like Coordinate \ $r_*$},
        title={Observer on a Constant $\rho$ Surface},
        legend pos = outer north east
     ]
	\addplot[samples = 500, domain = 0.01:6, thick, red]{ln(sinh(x))};
	\addplot[samples = 500, domain = 0.01:6, thick, blue]{-ln(sinh(x))};
	\addplot[samples = 500, domain = 0.01:6, thick, black]{x};
	\addplot[samples = 500, domain = 0.01:6, thick, red]{ln(sinh(x)) + 2};
	\addplot[samples = 500, domain = 0.01:6, thick, blue]{-ln(sinh(x)) + 2};
	\addplot[samples = 500, domain = 0.01:6, thick, red]{ln(sinh(x)) + 4};
	\addplot[samples = 500, domain = 0.01:6, thick, blue]{-ln(sinh(x)) + 4};
	\addplot[samples = 500, domain = 0.01:6, thick, black]{-x};
	\addplot[samples = 500, domain = 0.01:6, thick, black]{x+2};
	\addplot[samples = 500, domain = 0.01:6, thick, black]{-x+2};
	\addplot[samples = 500, domain = 0.01:6, thick, black]{x+4};
	\addplot[samples = 500, domain = 0.01:6, thick, black]{-x+4};
	\legend{Constant $\rho$ Observer, Constant $\rho$ (time reversed), Radial Null Geodesics}
	\end{axis}
\end{tikzpicture}
\captionof{figure}{Above the motion of an observer on a constant $\rho$ surface has been plotted in red for various values of $\rho$. The time reversed process is in blue and black lines correspond to ingoing and outgoing radial null geodesics.}\label{name}
\end{center}

From above, one can see that an observer on a surface of constant $\rho$ corresponds to radial infall into the black hole. The time reversed process has also been plotted and corresponds to an expanding surface. Considering this observer experiences a constant radial acceleration proportional to the cosmological constant, this suggests that the form of the metric (2.3) is well-adapted to infalling observers. 


\section{Conclusion}
The derivation here of the entropy of the non-rotating BTZ black hole differs from previous dual conformal descriptions in that the analysis holds at any spatial location. The same central charges originally computed by Brown and Henneaux, as well as the same conformal weight computed by Strominger, have been recovered without taking either a near horizon or $r \rightarrow \infty$ limit. This suggests that the degrees of freedom of 2+1 dimensional black holes are imprinted at any location around the black hole, which, as stated earlier, is physically appealing because in 2+1 dimensions there are no propagating degrees of freedom. Extending the current work to the full rotating BTZ black hole is currently in progress. 

One might also ask what relevance, if any, the above analysis has for more realistic higher dimensional black holes. In particular, since in higher dimensions there are propagating degrees of freedom, one does not necessarily expect there to exist a dual conformal description at any location. There is hope, however, that the boundary conditions here might help shed some light on the appropriate boundary conditions to impose when taking the near horizon limit in the higher dimensional case. Work in this direction is already underway and, if successful, would provide a more universal and systematic procedure for calculating the entropy of black holes from a dual conformal description. 

I would also like to mention that a related question has recently been considered by Comp\`ere et al. in the asymptotically de Sitter context \cite{DeSitterComp}, in which the central charges of $c^{\pm} = 3\ell/2G$ are also computed at any spatial location. There are, however, some key differences in the calculations. In particular, the boundary conditions appear to be different compared to the ones used here. Nevertheless, considering both derivations yield the expected central charges, it would be interesting to find some correspondence between them.

\subsection*{Acknowledgements} 
I would like to thank Steven Carlip for his ideas and insights without which this work would have neither been conceived nor completed. This work was supported in part by Department of Energy grant DE-FG02-91ER40674.


 \renewcommand{\theequation}{A.\arabic{equation}}
 \setcounter{equation}{0} 
 \section*{APPENDIX \ \ \ \ Relevant Aspects of Hamiltonian Gravity} 
Let us start with an arbitrary metric in an n-dimensional spacetime written in ADM form \cite{EffectiveConform}

\begin{equation}
	ds^2 = -N^2 dt^2 + q_{ij}(dx^i + N^i dt)(dx^j + N^j dt)
\end{equation}
where $N$ is the lapse, $N^i$ is the shift and $q_{ij}$ is the spatial metric. The momentum canonically conjugate to the spatial metric is 

\begin{equation}
	\pi^{ij} = \sqrt{q}(K^{ij} - q^{ij}K)
\end{equation}
where $K_{ij}$ is the extrinsic curvature of a constant time slice, $K$ is its trace and I have chosen units in which $16\pi G = 1$.  Units will sometimes be restored for clarity. 

In this formulation, symmetries are generated by the Hamiltonian

\begin{equation}
	H[\xi] = \int{d^{(n-1)}x (\xi^{\perp} \mathcal{H} + \hat{\xi}^i \mathcal{H}_i)}
\end{equation}
where $\mathcal{H}$ and $\mathcal{H}^i$ are, respectively, the Hamiltonian and momentum constraints:

\begin{equation}
	\mathcal{H} = \frac{1}{\sqrt{q}}\Big(\pi^{ij}\pi_{ij} - \frac{1}{n - 2}\pi^2\Big) - \sqrt{q}\Big({}^{(n-1)}R - 2\Lambda\Big), \ \ \ \ \ \mathcal{H}^i = -2D_j \pi^{ij}
\end{equation}
Here, $D_j$ is the spatial covariant derivative compatible with the spatial metric $q_{ij}$ and $^{(n - 1)}R$ is the spatial Ricci scalar. On a manifold without boundary, the canonical variables $(q_{ij}, \pi^{ij})$ have Poisson brackets with the generators

\begin{equation}
	\begin{split}
		\{H[\xi], q_{ij}\} =& -\frac{2}{\sqrt{q}} \xi^{\perp} \Big(\pi_{ij} - \frac{1}{n - 2} q_{ij} \pi \Big) - \Big(D_i \hat{\xi}_j + D_j \hat{\xi}_i\Big) \\
		\{H[\xi], \pi^{ij}\} =& \ \sqrt{q} \ \xi^{\perp} \Big({}^{(n - 1)}R^{ij} - \frac{1}{2} q^{ij} \Big({}^{(n-1)}R - 2\Lambda\Big)\Big) + \frac{2}{\sqrt{q}} \xi^{\perp} \Big(\pi^{ik}\pi_k^j - \frac{1}{n-2} \pi\pi^{ij}\Big) \\
		&-\frac{1}{2\sqrt{q}} \xi^{\perp} q^{ij}\Big(\pi^{k\ell}\pi_{k\ell} - \frac{1}{n-2} \pi^2\Big) - \sqrt{q}\Big(D^iD^j\xi^{\perp} - q^{ij} D_kD^k\xi^{\perp}\Big)\\
		&-D_k\Big(\hat{\xi}^k\pi^{ij}\Big) + \pi^{ik}D_k\hat{\xi}^j + \pi^{ik}D_k\hat{\xi}^i\\
	\end{split}
\end{equation}	
The above transformation is not exactly a full spacetime diffeomorphism but it is equivalent on-shell to a diffeomorphism generated by the vector field $\xi^{\mu}$ related  to the surface deformation parameters $(\xi^{\perp}, \hat{\xi}^i)$ by \cite{TeitSDParameter}

\begin{equation} 
	\xi^{\perp} = N\xi^t, \ \ \ \ \hat{\xi}^i = \xi^i + N^i \xi^t
\end{equation}
	
The Poisson brackets between two generators closes

\begin{equation}
	\{H[\xi], H[\eta]\} = H[\{\xi,\eta\}_{SD}]
\end{equation}
if ones takes as the Lie bracket of the surface deformations \cite{TeitSDParameter}

\begin{equation}
	\begin{split} 
		\{\xi, \eta\}^{\perp}_{SD} &= \hat{\xi}^i D_i \eta^{\perp} - \hat{\eta}^i D_i \xi^{\perp} \\
		\{\xi,\eta\}^i_{SD} &= \hat{\xi}^k D_k \eta^i - \hat{\eta}^k D_k \hat{\xi}^i + q^{i k} (\xi^{\perp} D_k \eta^{\perp} - \eta^{\perp} D_k \xi^{\perp}) \\
	\end{split}	
\end{equation}
The Lie bracket (A.8) assumes the surface deformation parameters are independent of the canonical variables $(q_{ij}, \pi^{ij})$. When this is not the case the surface deformation parameters can have nontrivial Poisson brackets with the generators. The algebra of the generators $H[\xi]$ closes if instead of (A.8) one uses the full surface deformation bracket \cite{BrownADS, TeitSDParameter}:

\begin{equation}
	\begin{split} 
		\{\xi, \eta\}_{full}^{\perp} &= \hat{\xi}^i D_i \eta^{\perp} - \hat{\eta}^i D_i \xi^{\perp} + \{H[\xi], \eta^{\perp}\}_{PB} - \{H[\eta], \xi^{\perp}\}_{PB}\\
		\{\xi,\eta\}_{full}^i &= \hat{\xi}^k D_k \eta^i - \hat{\eta}^k D_k \hat{\xi}^i + q^{i k} (\xi^{\perp} D_k \eta^{\perp} - \eta^{\perp} D_k \xi^{\perp}) + \{H[\xi], \eta^i\}_{PB} - \{H[\eta], \xi^i\}_{PB} \\
	\end{split}	
\end{equation}

On a manifold with boundary, there exist additional complications. In particular, the generators $H[\xi]$ are not generally differentiable \cite{ReggeAddBterm}. In order to obtain a well-defined variational principle, one must typically add a boundary term $B[\xi]$ to the generators. $B[\xi]$ must be chosen to cancel the boundary variation of the Hamiltonian \cite{HamBterms}

\begin{equation}
	\begin{split}
	\delta H[\xi] = ... - \frac{1}{16\pi G} \int_{\partial \Sigma}d^{n-2}x \Big\{ \sqrt{\sigma} &\Big[\xi^{\perp}(n^k \sigma^{\ell m} - n^m \sigma^{\ell k})D_m \delta q_{\ell k} \\
	&- D_m \xi^{\perp}(n^k \sigma^{\ell m} - n^m \sigma^{\ell k})q_{\ell k} \Big] + 2\hat{\xi}^i\delta \pi^n_i - \hat{\xi}^n \pi^{ij}\delta q_{ij}\Big\} 
	\end{split}	
\end{equation}
where $\sigma_{ij}$ is the induced metric on the boundary, $\sigma$ is its determinant and $n^k$ is the unit normal to the boundary.

The new generators $\widetilde{H}[\xi] = H[\xi] + B[\xi]$ have well-defined Poisson brackets by definition, which can be exploited as a way of isolating any possible central terms. Following the derivation performed by Carlip \cite{EffectiveConform}, one can start by writing

\begin{equation}
	\{ \widetilde{H} [\xi], \widetilde{H}[\eta]\}_{PB} = \widetilde{H}[\{\xi,\eta\}_{SD}] + K[\xi,\eta]
\end{equation}
where $K[\xi,\eta]$ represents a central term that may or may not be zero. Subtracting $\widetilde{H}[\{\xi,\eta\}_{SD}]$ from both sides, evaluating the functional derivatives from the Poisson Brackets and integrating by parts yields

\begin{equation}
\begin{split}
	K[\xi,\eta] = -B[\{\xi,\eta\}_{SD}] &- \frac{1}{8 \pi G} \int_{\partial \Sigma}{d^{n - 2}x \sqrt{\sigma} n^k \Big[ \frac{1}{\sqrt{q}}\pi_{ik} \{\xi,\eta\}^i_{SD} - \frac{1}{2\sqrt{q}} (\hat{\xi}_k \eta^{\perp} - \hat{\eta}_k \xi^{\perp}) \mathcal{H} } \\
	&+ (D_i \hat{\xi}_k D^i \eta^{\perp} - D_i \hat{\eta}_k D^i \xi^{\perp}) - (D_i \hat{\xi}^i D_k \eta^{\perp} - D_i \hat{\eta}^i D_k \xi^{\perp}) \\
	&+ \frac{1}{\sqrt{q}}(\hat{\eta}_k \pi^{mn} D_m \hat{\xi}_n - \hat{\xi}_k \pi^{mn} D_m \hat{\eta}_n) + (\xi^{\perp} \hat{\eta}^i - \eta^{\perp}\hat{\xi}^i)^{(n-1)}R_{ik}\Big]
\end{split}	
\end{equation}	
In general, due to the presence of $B[\{\xi,\eta\}_{SD}]$ above, one might think that the central term is sensitive to the specific boundary conditions. In some cases, however, this can be avoided. In particular, as discussed by Carlip, if one is able to find a Witt algebra $\{\xi, \eta\}_{SD} = \xi \eta' - \eta \xi'$ then the boundary term will depend only on this combination. A true central term, on the other hand, has a characteristic three-derivative structure $\xi'\eta'' - \eta'\xi''$. In this case, the boundary term in (A.12) is not expected to contribute to the presence of central charges.



\begin{thebibliography}{20}

\bibitem{Hawk}
S. W. Hawking, Nature 248, 30 (1974).

\bibitem{Beken}
J. D. Bekenstein, Phys. Rev. D7, 2333 (1973).

\bibitem{Strominger}
A. Strominger, JHEP 9802 (1998) 9802, arXiv:hep-th/9712251v3.

\bibitem{StringTheoryApproach}
D. Birmingham, I. Sachs, and S. Sen, Phys. Lett. B424 (1998) 275, arXiv:hep-th/9801019.

\bibitem{BrownADS}
J. D. Brown and M. Henneaux, Commun Math. Phys. 104 (1986) 207.

\bibitem{ConformOrigin}
S. Carlip, in \textit{Field Theory, Integrable Systems and Symmetries}, edited by F. Khanna and L. Vinet (Les Publications CRM, Montreal, 1997), arXiv:gr-qc/9509024.

\bibitem{BTZsolution}
M. Banados, C. Teitelboim, J. Zanelli, Phys. Rev. Lett. 69, 1849 (1992), arXiv:hep-th/9204099.

\bibitem{Cardy1}
J. A. Cardy, Nucl. Phys. B 270 (1986) 186.

\bibitem{Cardy2}
H. W. J. Blote, J. A. Cardy, and M. P. Nightingale, Phys. Rev. Lett. 56 (1986) 742.

\bibitem{CarlipBCHorizon}
S. Carlip, Phys. Rev. Lett. 82 (1999) 2828, arXiv:hep-th/9812013.

\bibitem{SolodukhinBCHorizon}
S. N. Solodukhin, Phys. Lett. B454, 213 (1999), arXiv:hep-th/9812056.

\bibitem{CarlipBCHorizon2}
S. Carlip, Class. Quant. Grav. 16 (1999) 3327, arXiv:gr-qc/9906126.

\bibitem{ADM}
R. Arnowitt, S. Deser, and C.W. Misner, in \textit{Gravitation: an Introduction to Current Research}, edited by L. Witten (Wiley, New York, 1962), arXiv:gr-qc/0405109.

\bibitem{TeitSDParameter}
C. Teitelboim, Ann. Phys. 79 (1973) 542.

\bibitem{ReggeAddBterm}
T. Regge, C. Teitelboim, Ann. Phys. 88 (1974) 286.

\bibitem{HamBterms}
J. D. Brown, S. R. Lau, and J. W. York, Ann. Phys. 297 (2002) 175, arXiv:gr-qc/0010024.

\bibitem{EffectiveConform}
S. Carlip, Entropy 13(7) (2011) 1355, arXiv:gr-qc/1107.2678.

\bibitem{Whatwedontknow}
S. Carlip, Class. Quant. Grav. 15 (1998) 3309, arXiv:hep-th/9806026.

\bibitem{UniqueC}
I. M. Gel'fand and D. B. Fuks, Func. Anal. Appl. 2 (1968) 342.

\bibitem{CFTyellowpages}
P. Di. Francesco, P. Mathieu, and D. S\'en\'echal, \textit{Conformal Field Theory} (Springer, 1997).

\bibitem{ADSVac}
O. Coussaert and M. Henneaux, Phys. Rev. Lett. 72 (1994) 622.

\bibitem{BoundaryDynamics}
M. Banados, T. Brotz, and M. E. Ortiz, Boundary Dynamics and the Statistical Mechanics of the 2+1 Dimensional Black Hole, hep-th/9802076.

\bibitem{Waldtext}
R. M. Wald, \textit{General Relativity} (University of Chicago Press, 1984)

\bibitem{Birmingham_CFT}
D. Birmingham, I. Sachs, and S. Sen, Phys. Lett. B413, 281 (1997), arXiv:hep-th/9707188

\bibitem{DeSitterComp}
G. Comp\`ere, L. Donnay, P.-H. Lambert, and W. Schulgin, JHEP 1503 (2015) 158, arXiv:hep-th/1411.7873

\end{thebibliography}
\end{document}